\documentclass[amsmath,longbibliography,reprint,aip, apl]{revtex4-1}
\usepackage{hyperref}
\usepackage{graphicx}
\begin{document}
\title{Spectral tuning of optical coupling between air-mode nanobeam cavities and individual carbon nanotubes}
\author{H.~Machiya}
\affiliation{Nanoscale Quantum Photonics Laboratory, RIKEN, Saitama, 351-0198, Japan}
\affiliation{Department of Electrical Engineering, The University of Tokyo, Tokyo, 113-8656, Japan}
\author{T.~Uda}
\affiliation{Nanoscale Quantum Photonics Laboratory, RIKEN, Saitama, 351-0198, Japan}
\affiliation{Department of Applied Physics, The University of Tokyo, Tokyo, 113-8656, Japan}
\author{A.~Ishii}
\affiliation{Nanoscale Quantum Photonics Laboratory, RIKEN, Saitama, 351-0198, Japan}
\affiliation{Quantum Optoelectronics Research Team, RIKEN Center for Advanced Photonics, Saitama, 351-0198, Japan}
\author{Y.~K.~Kato}
\email[Corresponding author: ]{yuichiro.kato@riken.jp}
\affiliation{Nanoscale Quantum Photonics Laboratory, RIKEN, Saitama, 351-0198, Japan}
\affiliation{Quantum Optoelectronics Research Team, RIKEN Center for Advanced Photonics, Saitama, 351-0198, Japan}
\begin{abstract}
We demonstrate control over optical coupling between air-suspended carbon nanotubes and air-mode nanobeam cavities by spectral tuning.
Taking advantage of the large dielectric screening effects caused by adsorbed molecules, laser heating is used to blueshift the nanotube photoluminescence.
Significant increase of the cavity peak is observed when the nanotube emission is brought into resonance, and the spontaneous emission enhancement is estimated from the photoluminescence spectra.
We find that the enhancement shows good correlation to the spectral overlap of the nanotube emission and the cavity peak.
Our technique offers a convenient method for controlling the optical coupling of air-suspended nanotubes to photonic structures.
\end{abstract}
\keywords{carbon nanotubes, photoluminescence, cavity}

\maketitle

Semiconducting carbon nanotubes (CNTs) are an attractive material for integrated photonics owing to its exceptionally small size,\cite{Pyatkov2016} and CNT-based discrete components such as electrically driven emitters,\cite{Mueller2010,Higashide2017} photodetectors,\cite{Lee2005,Kumamoto2014} and single photon sources \cite{Jeantet2016,Ishii2017} have been demonstrated.
Optical coupling to photonic structures \cite{Watahiki2012a,Imamura2013,Miura2014a,Pyatkov2016,Noury2015,Graf2016} is one of the next key requirements for on-chip integration.
Cavities, in particular, offer quantum electrodynamical effects for increased performance and functionality by confining electric fields into a small mode volume with a low optical loss.
Exciton polaritons can form when nanotube emission is strongly coupled to a cavity, \cite{Graf2016} whereas interaction of CNTs with cavities results in an increased spontaneous emission rate in the weak coupling regime.\cite{Watahiki2012a,Imamura2013,Miura2014a,Noury2015,Pyatkov2016}
The spectral overlap of the emission with the cavity is an important factor in determining these effects, and the overlap can be changed by preparing samples with different cavity lengths for coupling to ensembles of CNTs.\cite{Graf2016}
Sophisticated fiber microcavities enable {\it in-situ} tunable coupling to individual CNTs,\cite{Jeantet2016} but it still remains a challenge to perform post-fabrication tuning for on-chip cavities.

In this work, we demonstrate spectral tuning of air-suspended carbon nanotubes coupled to integrated nanobeam cavities.
We utilize air-mode cavities \cite{Miura2014a,Liang2015,Pyatkov2016} with large electric fields in air for efficient coupling to suspended CNTs.
Laser heating is used to desorb molecules from the nanotubes, which offers a simple method for adjusting the nanotube emission through reduction of dielectric screening.
We are able to modify the detuning by as much as 25~meV, inducing a drastic change in the cavity peak intensity.
Cavity quantum electrodynamics model is used to analyze the data, and we find that the enhancement of the cavity mode can mostly be explained by the spectral overlap of the nanotube emission with the cavity peak.

\begin{figure}
\includegraphics{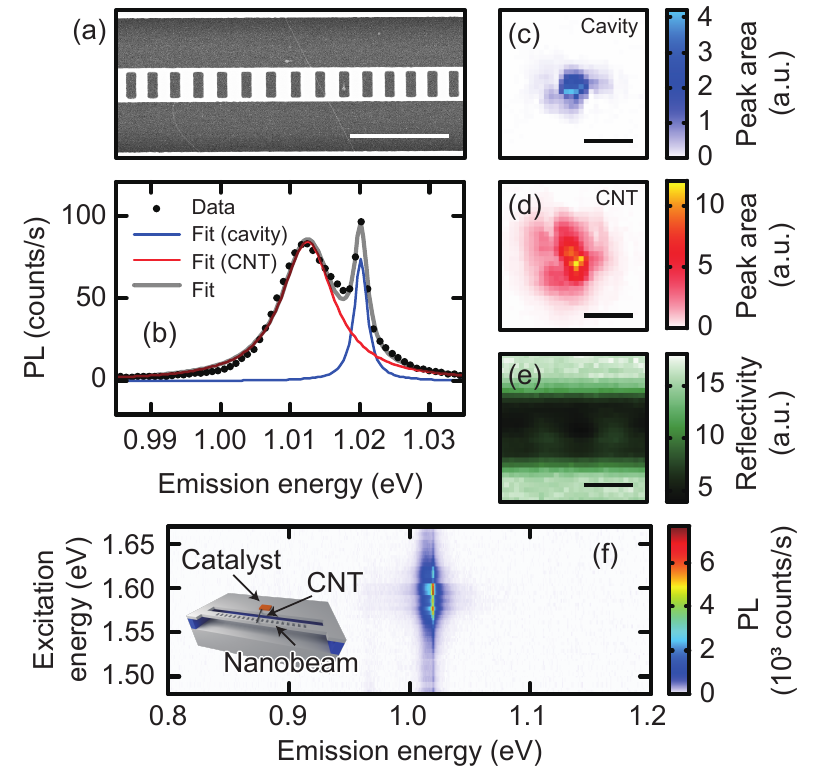}
\caption{\label{fig1}(a) Scanning electron microscope image of a fabricated cavity with an air-suspended CNT. The scale bar is 2~$\mu$m.
(b) Typical PL spectrum of a cavity coupled to a CNT. Black dots are data and solid lines are the Lorentzian multi-peak fits. $P=1~\mu$W and an excitation laser energy of 1.594~eV are used. 
(c) and (d) Spectrally-resolved PL images of the cavity emission and the nanotube PL, respectively. 
PL spectra are fitted to two Lorentzian functions with fixed center energies and linewidths corresponding to the cavity and the CNT peaks. The peak areas are used to construct the images.
(e) Reflectivity image taken simultaneously with (c) and (d).
For (c)-(e), $P=200~\mu$W and an excitation energy of 1.590~eV are used, and the scale bars are 2~$\mu$m.
(f) PLE map of the device taken with $P=50~\mu$W. Inset is a schematic of a device. 
In (b) and (f), spectra are taken where the cavity emission is maximized.
For all PL measurements, laser polarization is perpendicular to the nanobeam.}
\end{figure}

The air-mode nanobeam cavities \cite{Miura2014a} [Fig.~\ref{fig1}(f) inset] are fabricated from silicon-on-insulator wafers diced into 20-mm square chips.
Electron beam lithography defines the cavity pattern and the 260-nm-thick top silicon layer is etched through by a Bosch process in an inductively coupled plasma etcher.
The 1-$\mu$m-thick buried oxide layer is then removed by 20\% hydrofluoric acid, followed by another electron beam lithography process which defines the catalyst area.
We finally spin-coat Fe/silica catalyst, and grow CNTs onto the cavity by alcohol chemical vapor deposition at 800$^\circ$C.\cite{Maruyama2002a}
A scanning electron microscope image of a typical device is shown in Fig.~\ref{fig1}(a).

The devices are measured using a room-temperature confocal microscope system.\cite{Ishii2015a}
An output of a wavelength-tunable Ti:sapphire laser is focused onto the sample using an objective lens with a numerical aperture of 0.8 and a working distance of 3.4~mm.
We mount the sample on a motorized three-dimensional feedback stage, which allows for efficient automated measurements.
Photoluminescence (PL) is collected using the same objective lens and detected using a liquid-nitrogen-cooled InGaAs diode array attached to a spectrometer.
Laser reflection comes back through the objective lens and is detected with a Si photodiode, which is used to align to the cavity position.
Samples are placed in nitrogen to avoid oxidation of nanotubes at high excitation powers.

The cavities are designed to have resonances near the emission energy of chiralities with large population to maximize the probability of coupling, and more than 10000 cavities are prepared on a chip.
Nanotubes optically coupled to the cavity are located by one-dimensional PL scans over the nanobeams.
The automated scan results are filtered by performing peak detection and linewidth estimation.
Since the cavity linewidth is considerably narrower than the nanotube linewidth at room temperature, devices showing a sharp peak component are chosen for detailed characterization.

A typical PL spectrum of a single CNT coupled to a cavity is shown in Fig.~\ref{fig1}(b) where the excitation laser power $P=1~\mu$W is used.
We use two Lorentzian functions to decompose the spectrum into the sharp cavity mode (blue) and the broad nanotube emission (red), and spectrally resolved PL images [Figs.~\ref{fig1}(c) and (d)] are obtained by spatially mapping each peak area.
Combined with the reflectivity image [Fig.~\ref{fig1}(e)], we find that the sharp component is localized at the center of the device [Fig.~\ref{fig1}(c)], consistent with the spatial extent of the cavity mode.
In comparison, the spectrally broad component is distributed across the width of the trench [Fig.~\ref{fig1}(d)], showing the location and the length of the CNT.
To confirm that the CNT is individual, PL excitation map is also taken [Fig.~\ref{fig1}(f)].
A single peak in the map confirms that the cavity is coupled to an individual CNT, and its chirality is assigned to be (10,5) by comparing to tabulated data.\cite{Ishii2015a}

\begin{figure}
\includegraphics{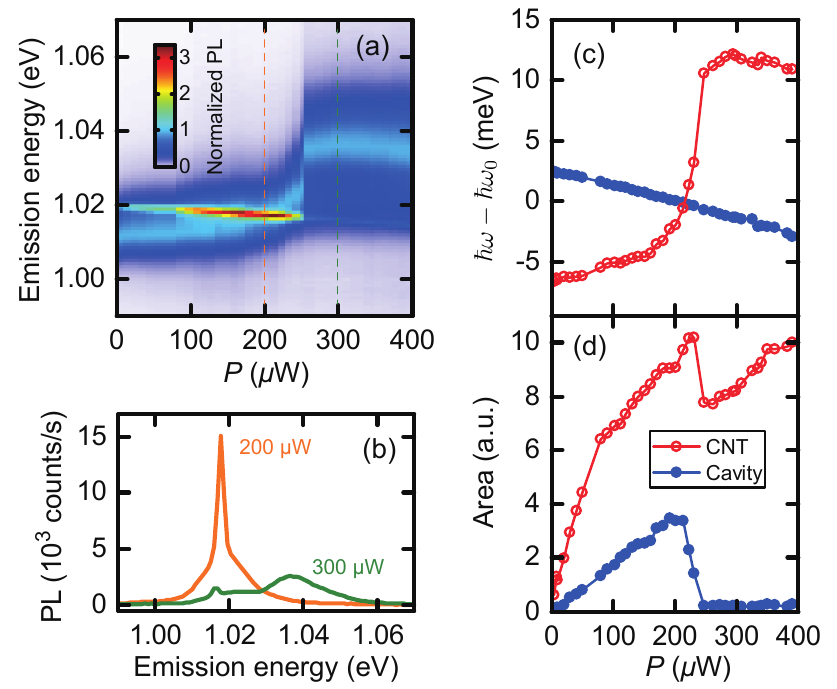}
\caption{\label{fig2}(a) Excitation power dependence of PL. PL is normalized with respect to the nanotube peak height. Vertical dashed lines indicate the powers where PL spectra in (b) are taken. (b) Comparison of PL spectra taken at $P=200~\mu$W (orange) and $300~\mu$W (green). 
(c) and (d) Excitation power dependence of the peak center energy and peak area, respectively. 
Red open circles are obtained from the CNT peak, and the blue dots are taken from the cavity emission.
The peak center energies are plotted as the difference from the resonant energy $\hbar\omega_0=1.018~$eV.
For (a)-(d), excitation laser energy is tuned to $E_{22}$ resonance and laser polarization is perpendicular to the nanobeam.}
\end{figure}

Looking at Fig.~\ref{fig1}(b), the nanotube emission needs blueshifting by $\sim10~$meV to achieve resonance.
In order to tune the optical transition energy, 
we take advantage of the sensitivity of air-suspended nanotubes to dielectric screening.
By controlling the amount of adsorbed molecules through temperature or pressure, large spectral shifts of up to 30~meV have been observed.\cite{Finnie2005,Lefebvre2008,Homma2013,Uda2017}
We utilize laser heating to induce molecular desorption in our devices, which is a simple and convenient method.
As the excitation power is increased, the nanotube peak blueshifts due to molecular desorption while the cavity peak gradually brightens [Fig.~\ref{fig2}(a)].
The nanotube emission is brought into resonance at $P\sim200~\mu$W, enhancing cavity emission intensity.
By further increasing the excitation power, the cavity peak becomes weaker again as the nanotube peak continues to blueshift.
Above $P\sim250~\mu$W, we do not find additional blueshifting because the molecules are fully desorbed.

In Fig.~\ref{fig2}(b), we compare on- and off-resonance spectra, which are indicated by orange and green curves, respectively.
We find an enhancement of the cavity peak height by a factor of 10 at resonance, while the nanotube peak height remains comparable.
The significant difference in the cavity peak height indicates that the optical coupling to the nanotube emission is modified by the spectral shift.
The spectral lineshape of the nanotube peak is changed at high powers (green curve), probably due to the inhomogeneous molecular desorption.

To quantitatively characterize the spectral tuning effect, the excitation power dependence is analyzed by decomposing the spectra into the cavity and the nanotube peak components.
We first fit the sharp cavity peak to a single Lorentzian function using a fitting window of  $\sim$10~meV, yielding the cavity peak energy $\hbar\omega_{\mathrm{cav}}$, linewidth $\gamma$, and peak area $I_{\mathrm{cav}}$.
Due to the complex nanotube lineshape, we let the nanotube peak spectrum $L_{\mathrm{CNT}}(\omega)$ to be the fitting residual, where $\omega$ is the frequency.
The nanotube peak area $I_{\mathrm{CNT}}$ is defined as the integral
\begin{equation}
I_{\mathrm{CNT}}=\int L_{\mathrm{CNT}}(\omega)\mathrm{d}\omega,
\end{equation}
while the peak energy $\hbar\omega_{\mathrm{CNT}}$ is defined as the weighted average
\begin{equation}
\hbar\omega_{\mathrm{CNT}}=\hbar\frac{\int \omega L_{\mathrm{CNT}}(\omega)\mathrm{d}\omega}{\int L_{\mathrm{CNT}}\mathrm{d}\omega}.
\end{equation}

In Fig.~\ref{fig2}(c), we plot the peak energies as a function of the excitation power, and blueshifting of the nanotube peak by as much as 18~meV (red open circles) is observed.
A slight redshift of the cavity peak (blue dots) likely originates from the heating-induced increase of refractive index.\cite{ElKurdi2008}
The energy shifts of the nanotube and cavity together allow for energy detuning $\hbar\omega_{\mathrm{cav}}-\hbar\omega_{\mathrm{CNT}}$ to be changed from $+9~$meV to $-14$~meV.
The tuning range is sufficiently large for modulating the optical coupling of the nanotubes to the cavity, as it considerably exceeds the nanotube emission linewidth at room temperature.
We note that the shifts are reversible upon reduction of the excitation power.

We now turn our attention to the excitation power dependence of emission intensities [Fig.~\ref{fig2}(d)].
$I_{\mathrm{CNT}}$ increases linearly at low excitation powers and then becomes sublinear as observed previously,\cite{Ishii2015a} with a drastic drop at $P=250~\mu$W [Fig.~\ref{fig2}(d), red open circles].
Since the $E_{22}$ resonance also blueshifts by molecular desorption,\cite{Lefebvre2008,Homma2013,Uda2017} the sudden drop can be understood by the shifting of the $E_{22}$ resonance away from the excitation energy.
In comparison, we observe a linear increase of $I_{\mathrm{cav}}$ with excitation power, for the region below $P=190~\mu$W [Fig.~\ref{fig2}(d), blue dots].
The sublinearity of $I_{\mathrm{CNT}}$ is likely cancelled out by the improvement of coupling, because the nanotube peak approaches resonance in this power region.
The cavity peak area then rapidly reduces above $P=200~\mu$W as a result of the increased energy detuning.

Since we are changing the excitation power, we need to characterize the optical coupling by evaluating emission enhancement.
Assuming that stimulated emission is negligible, the spontaneous emission enhancement factor is given by
\begin{equation}
F=\frac{I_{\mathrm{cav}}/\eta_{\mathrm{cav}}}{I_{\mathrm{CNT}}/\eta_{\mathrm{CNT}}},
\end{equation}
where $\eta_{\mathrm{CNT}}$ and $\eta_{\mathrm{cav}}$ are the collection efficiencies of the nanotube and the cavity emission, respectively.
To obtain conservative estimates of $F$, $\eta_{\mathrm{CNT}}/\eta_{\mathrm{cav}}=0.49$ is used.\cite{Miura2014a}

\begin{figure}
\includegraphics{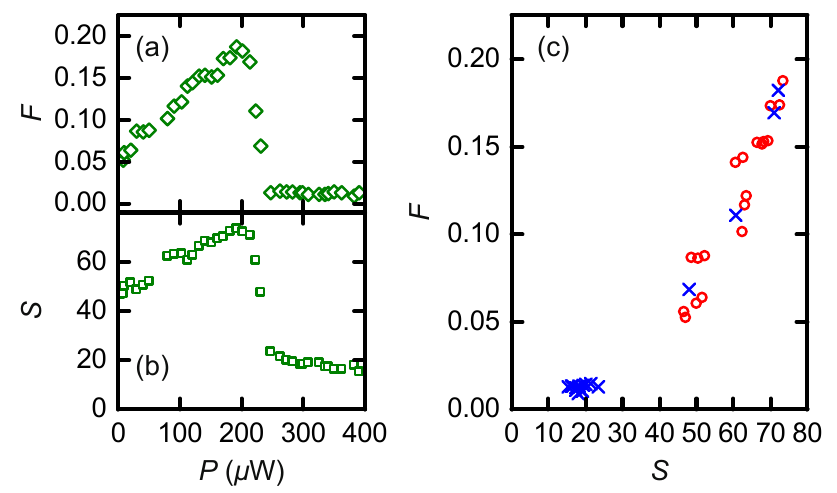}
\caption{\label{fig3} (a) and (b) Excitation power dependence of $F$ and $S$, respectively. $F$ and $S$ are calculated from the results in Fig.~\ref{fig2}. (c) $F$ as a function of $S$. Open circles indicate red-detuned nanotube emission ($P\leq190~\mu$W) and crosses correspond to blue-detuned conditions ($P>190~\mu$W). }
\end{figure}

We have calculated $F$ using the data shown in Fig.~\ref{fig2} and the results are plotted in Fig.~\ref{fig3}(a).
As the excitation power is increased up to $P=190~\mu$W, the enhancement factor increases from 0.056 to 0.19, and $F$ rapidly decreases by a factor of 15 at $P=250~\mu$W.
The large modification to the enhancement factor is consistent with the spectral shifts in this range.
Above $P=250~\mu$W, we do not observe much excitation power dependence.

The enhancement factor is known to be determined by the spectral overlap, spatial overlap, and polarization matching.\cite{Noda2007,Unitt2005}
If we assume that the nanotube does not move during our experiment, $F$ should only depend on the spectral overlap.
As the nanotube lineshape is complex, we need to evaluate the spectral overlap without assuming a particular emitter lineshape.
We first consider a simple case of a monochromatic emitter coupled to a cavity with a Lorentzian lineshape.
The spontaneous emission enhancement at a frequency $\omega$ is then given by
\begin{equation}
\frac{3}{4\pi^2}\left(\frac{\lambda_0^3}{V}\right)\left(\frac{\omega}{\gamma}\right)\frac{(\gamma/2)^2}{(\omega-\omega_{\mathrm{cav}})^2+(\gamma/2)^2}
\end{equation}
where $\lambda_0$ and $V$  are the wavelength and the mode volume of the cavity, respectively.\cite{VanExter1996}
We multiply the cavity enhancement and the normalized lineshape of the nanotube emission, and integrate over the full spectrum to obtain
\begin{equation}
\label{eq.f}
F=\frac{3}{4\pi^2}\left(\frac{\lambda_0^3}{V}\right)S
\end{equation}
where the integral
\begin{equation}
\label{eq.s}
S=\int\left(\frac{\omega}{\gamma}\right)\frac{(\gamma/2)^2}{(\omega-\omega_{\mathrm{cav}})^2+(\gamma/2)^2} \frac{L_{\mathrm{CNT}}(\omega)}{I_{\mathrm{CNT}}}\mathrm{d}\omega
\end{equation}
is the generalized form of the spectral overlap.
$1/S$ reduces to $1/Q_{\mathrm{cav}}+1/Q_{\mathrm{CNT}}$ under an assumption that the nanotube lineshape is a Lorentzian resonant with the cavity,\cite{VanExter1996} where $Q_{\mathrm{cav}}$ and $Q_{\mathrm{CNT}}$ are the quality factors of the cavity and the nanotube peaks, respectively.
If $Q_{\mathrm{cav}}$ is considerably larger than $Q_{\mathrm{CNT}}$, the upper bound of $S$ would be given by $Q_{\mathrm{CNT}}$.

The spectral overlap can be calculated for arbitrary lineshapes using Eq.~(\ref{eq.s}), and results of numerical integration are plotted in Fig.~\ref{fig3}(b).
When the excitation power is increased, we observe a crossover from a gradual increase to a rapid decrease at $P=190~\mu$W as the nanotube peak shifts through the cavity peak.
The maximum spectral overlap $S=73$ is reasonable, considering that $Q_{\mathrm{CNT}}\sim80$.

Correlation between the enhancement factor and the spectral overlap is examined in Fig.~\ref{fig3}(c).
Indeed, $F$ increases monotonically with $S$ as expected from Eq.~(\ref{eq.f}).
We note that the detuning changes sign above and below $P=190~\mu$W, but similar enhancements are obtained for red- (open circles) and blue-detuned (crosses) nanotube emission.
It is reasonable that detuning sign does not have an effect, because $F$ should only depend on $S$.

\begin{figure}
\includegraphics{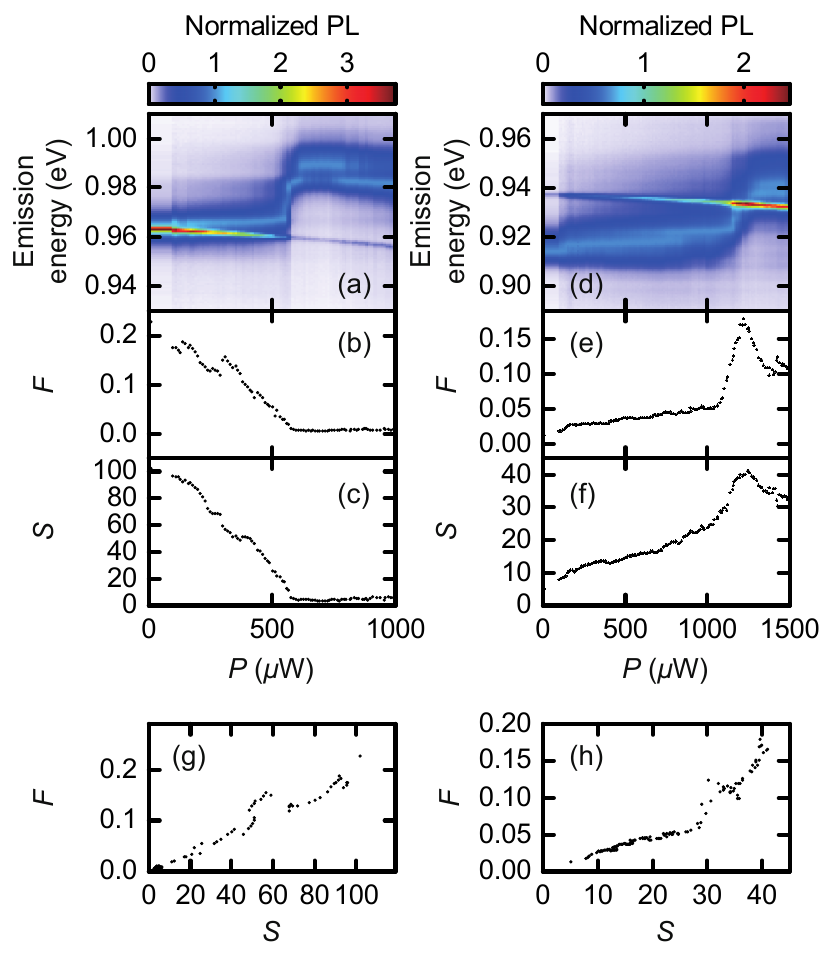}
\caption{\label{fig4}
(a) Normalized PL, (b)  $F$, and (c) $S$ as a function of excitation power for the second device.
(d) Normalized PL, (e)  $F$, and (f) $S$ as a function of excitation power for the third device.
(g) and (h) $S$ dependence of $F$ for the second and the third devices, respectively. 
Excitation laser energy and polarization are tuned to maximize the signal.
}
\end{figure}

In Fig.~\ref{fig4}, we present two additional devices that have different initial detunings.
In the second device, the nanotube peak is initially resonant with the cavity at the lowest excitation power [Fig.~\ref{fig4}(a)].
When the excitation power is increased, molecular desorption shifts the nanotube peak away from the cavity, reducing the enhancement [Fig.~\ref{fig4}(b)].
The blueshifting results in the spectral overlap to decrease as well [Fig.~\ref{fig4}(c)], explaining the behavior of $F$.
The third device has an initial detuning even larger than the first device [Fig.~\ref{fig4}(d)].
Even for the fully desorbed state, there exists considerable emission enhancement [Fig.~\ref{fig4}(e)].
The spectral overlap increases with the excitation power, but shows a less pronounced increase at the resonance compared to $F$ [Fig.~\ref{fig4}(f)].

Figs.~\ref{fig4}(g) and (h) are the $F$-$S$ plots of the second and the third devices, respectively.
We observe a monotonic increase of $F$ as observed in the first device, consistent with the understanding that the enhancement factor is determined by the spectral overlap.
It is interesting to note that slope of the $F$-$S$ plot should be $\frac{3}{4\pi^2}\frac{\lambda_0^3}{V}=3.2$ for perfect spatial and polarization overlaps.\cite{Miura2014a}
The linear fits to our experimental data, in comparison, yield slopes on the order of $10^{-3}$, which suggests there remains a lot of room for improving the coupling.
Better spatial and polarization alignment could be possible by site-controlled transfer of nanotubes or photonic crystal structures by micromanipulation.\cite{Wu2010,Tajiri2015}
We also note that the experimental slope may be an underestimate because of our conservative calculation of $F$.
The actual $F$ is expected to be higher since the collection efficiency of the nanotube emission should be enhanced due to the photonic crystal structure.\cite{Noda2007,Miura2014a}

It is intriguing that we observe a slight superlinearity in the $F$-$S$ plots for the first and the third devices [Figs.~\ref{fig3}(c) and \ref{fig4}(h)]. 
One possible explanation is the cavity-resonant optical forces.
We assume in our analysis that the spatial overlap does not change during our measurements, but the overlap may improve at resonance if the CNTs are pulled towards the electric field maximum.\cite{Ajiki2009,Mandal2010}
Another possibility is stimulated emission which contributes to additional emission into the cavity mode.
In the first and the third devices, the spectral overlap becomes larger at high excitation powers, which is preferable for stimulated emission.
The second device, in comparison, has smaller spectral overlap at high powers, which may explain why the superlinearity cannot be observed.
By controlling the pressure or temperature, it should be possible to keep the resonant condition and perform excitation power dependence measurements as well as photon correlation measurements \cite{Ota2017a} to identify the contribution from stimulated emission.

In summary, spectral tuning of air-suspended carbon nanotubes coupled to air-mode nanobeam cavities is demonstrated by laser-heating-induced molecular desorption.
We have performed excitation power dependence measurements and have shown that such molecular-scale effects provide a convenient {\it in-situ} method for controlling the coupling of as-grown CNTs to photonic structures.
The excitation power dependence is characterized using the enhancement factor, and we find that the numerically calculated spectral overlap can explain the behavior.
The tunability of optical coupling would be a key not only for fundamental understanding of light-matter interaction in CNTs but also for reconfigurable optical devices in the telecom band on a silicon platform.

We thank S. Chiashi and S. Maruyama for the use of the electron microscope.
This work is supported by JSPS (KAKENHI JP16H05962, JP16K13613), and MEXT (Photon Frontier Network Program, Nanotechnology Platform).
The samples were fabricated at the Center for Nano Lithography \& Analysis at The University of Tokyo.
H.M. is supported by RIKEN Junior Research Associate Program, and T.U. is supported by ALPS and JSPS Research Fellowship.

%\bibliography{reference}

\end{document}